\documentstyle[preprint,pre,aps,epsf,graphicx]{revtex}
\begin{document}

\draft

\title{On a variational principle model for the Nuclear Caloric curve}
 
\author{S. Das Gupta}

\address{Physics Department, McGill University, 
Montr{\'e}al, Canada H3A 2T8}

\date{\today}

\maketitle

\begin{abstract}
Following the lead of a recent work we perform a variational
principle model calculation for the nuclear caloric curve.  A Skyrme type
interaction with and without momentum dependence is used.
The calculation is done for a large nucleus, i.e, in the nuclear
matter limit.  Thus we address the issue of volume fragmentation
only.  Nonetheless the results are similar to the previous,
largely phenomenological calculation for a finite nucleus.
We find the onset of fragmentation can be sudden as a function
of temperature/excitation energy.

\end{abstract}


\section{Introduction}
At low excitation energy the nucleus deexcites by binary sequential decay.
This is envisaged by the compound nucleus hypothesis.  It is usually believed
that beyond a certain excitation, binary decay gives way to simultaneous 
break up \cite{Dasgupta1}.  This break up may well depend upon the
details of collision, i.e., whether the energy was pumped in by heavy
ion collisions or say, by a proton hitting a nucleus.  For multifragmentation,
it is conjectured that the nucleus expands to three/four times the normal
volume and breaks up into various fragments, dictated by phase space.
Calculations based on this picture have been successful in fitting many
data.  The SMM (statistical multifragmentation model) uses these ideas
and predicted (well before any experiments) that one should see a maximum
in the specific heat at temperatures around 5 MeV \cite{Bondorf1}.  This
was indeed found \cite{Pochodzalla}.  The appearence of a maximum in
the specific heat led to a very interesting speculation that this
may be a signature of phase transition.  This is more dramatically
demonstrated in the canoncal thermodynamic model (which is in the same
spirit as the SMM but is much easier to implement).  Here at low
temperature there is a large blob of matter which breaks up into many 
pieces and the maximum in specific heat is obtained in this 
region.  The thermodynamic model can be extrapolated to large numbers
of particles.  In this limit it is shown that the break up is very 
sudden as a function of temperature.  This is a case of first order
phase transition and the maximum of specific heat is obtained at the
phase transition \cite{Dasgupta2}.

The beginning of such models started with Bevalac energy collisions
where it was natural to assume that the system formed by collisions
is first compressed and then begins to decompress.  During the
decompression stage, the products in terms of mass numbers, velocities
etc. will evolve finally reaching a just streaming stage.  While the 
changes will be continuous one can probably get an adequate
description by assuming that all the changes can be described by
assuming a constant freeze-out volume and doing a phase-space
calculation (thermodynamics) at this volume.  One can use the 
Boltzmann-Uehling-Uhlenbeck microscopic simulations \cite{Bertsch}
in support of an initial compression, followed by expansion and
finally free streaming.

It is of course clear that a calculation of the caloric curve based
on such models is not, in any simple way, connected to a variational
principle.  The present work is inspired by a recent calculation which showed
that, in a different model based on a variational principle,
a plateau in the caloric curve (hence
a maximum in specific heat) can be reached in mononuclear configurations
as well \cite{Sobotka1}.  In this model, the nucleus expands when
excitation energy is pumped in.  Adopt a reasonable density profile
for the ground state and assume that when energy is pumped into the
nucleus, it expands in a self-similar fashion, i.e., $\rho(r,c)=
c^3\rho(rc)$ where $c$ is a function of the excitation energy $e^*$.
This excitation energy $e^*$ consists of two parts: a thermal part
and a compression part which, as a function of density,
increases quadratically about the ground state density \cite{Toke}.
At a given excitation energy, the nucleus expands till it reaches the 
maximum entropy (this is the variational principle of the calculation).  
Effects of interaction
on entropy is taken through a parametrisation of $m^*/m$ where
$m^*$ is the effective mass.  
A temperature is defined microcanonically.  The authors then find
that when they plot temperature against excitation energy, a plateau
is found around 5 to 6 MeV.

Our investigation addresses the question of mononuclear configuration
but with the following restriction .  There is no surface in
our calculation thus we are investigating break up as a volume
effect.  Stated differently, our calculation refers to nuclear
matter.  This restriction was dictated by our desire to do
a more microscopic calculation.  In \cite{Sobotka1} excitation
energy is fixed.  We do calculations first with fixed temperatures
(this is a more well-known practice in nuclear physics) but then
also do calculations with fixed energies.  In both the approaches,
the nucleus first expands as the temperature increases,
a plateau is reached but then the nucleus breaks up quite suddenly.
Once the nucleus breaks up, the model gives no guidance how the
caloric curve is to be calculated further, into higher energy.  
In order to continue one needs to formulate a different model.

\section{Temparure dependent mean field theory}
A familiar model in nuclear physics which allows study of 
the caloric curve in mononuclear configurations is the temparature
dependent mean field model (Hartee-Fock and/or Thomas-Fermi model).
This has certain advantages.  When one does a standard mean field 
calculation at a fixed temparture, one minimises the free energy
$F=E-TS$ \cite{March}.  This means that when we get the self-consistent
solution at a given temperature, we have obtained a solution which
has zero pressure.  If we draw a caloric curve with energies of these
solutions this caloric curve pertains to zero pressure.  The specific heat
that we will get will be $c_p$ with $p=0$.

Investigation of the caloric curve with temperature-dependent Thomas-Fermi
theory was done in the past \cite{De}.  For nuclei $^{150}$Sm and
$^{85}$Kr caloric curves were drawn and
a maximum in specific heat at temperature $\approx$10 MeV was found.  But 
there is ambiguity whether the systems stay mononuclear or not.  
The cause of the ambiguity is this.  In finite temperature mean field
theory (whether Thomas-Fermi or Hartree-Fock) single
particle states in the continuum are admixed through the finite 
temperature occupation factors.  These calculations are done in a finite
box and because of this admixture to the continuum,
the resulting density distributions become strongly dependent on the
volume of the box.  Thus in Fig. 2. of \cite{De} at temperature 10 MeV
if the calculation
for $^{150}$Sm is done by enclosing in a volume which is 4 times
the normal nuclear volume 
one still has a compact system with a central density
and a surface.  One would be tempted to call this a mononuclear system.
The density profile is totally different if the same
calculation is done where the confining volume is 8 times the
normal volume.  Now the density is smeared out almost uniformly
in the confining volume which would be the characteristics of a 
non-interacting gas.  Although the densities alter a great deal,
the maximum in the specific heat does not 
change much (close to 10 MeV in both the cases) but the widths of the
specific heat do.

\section{Volume break up in finite temperature mean field theory}
We can study the volume break up by working in the nuclear matter limit.
This is pursued in this work.  For a fixed temperature we 
do calculations for different densities.  The density where the free energy
per particle is minimised is the solution for this temperature; $e^*$
of this solution is the appropriate $e^*$ for this temperature. 
As expected, starting from zero temperature, the system expands.
The minima of free energy drop to lower and lower density as the
temperature increases.  But beyond a certain temperature, the minimum in free 
energy disappears.  The nucleus will now break apart.  This happens
just after $T$ flattens out as a function of $e^*$.  This is shown in figs. 1
and 2.

We give some details of the calculation.  We use the 
mometum dependent mean field of \cite{Welke,Gale1}.  The potential
energy density is given by
\begin{eqnarray}
v(\rho)=\frac{A}{2}\frac{\rho^2}{\rho_0}+\frac{B}{\sigma+1}\frac{\rho^{\sigma+
1}}{\rho_0^{\sigma}}+\frac{C}{\rho_0}\int\int d^3pd^3p'\frac{f(\vec r,\vec p)
f(\vec r,\vec p')}{1+[\frac{\vec p-\vec p'}{\Lambda}]^2}
\end{eqnarray}
Here $f(\vec r,\vec p)$ is phase-space density.  In nuclear matter, at 
zero temperature $f(\vec r,\vec p)=\frac{4}{h^3}\Theta (p_F-p)$
where 4 takes care of spin-isospin degeneracy.  At finite temperature
the theta function $\Theta(p_F-p)$ is replaced by Fermi occupation
factor (see details below). The potential felt by a particle is
\begin{eqnarray}
u(\rho,\vec p)=A[\frac{\rho}{\rho_0}]+B[\frac{\rho}{\rho_0}]^{\sigma}
+2\frac{C}{\rho_0}\int d^3p'\frac{f(\vec r,\vec p')}{1+[\frac{\vec p-
\vec p'}{\Lambda}]^2}
\end{eqnarray}
Here $A=$-110.44 Mev, $B$=140.9 MeV, $C$=-64.95 MeV, $\rho_0=0.16 fm^{-3}$,
$\sigma$=1.24 and $\Lambda=1.58p_F^0$.  This gives in nuclear matter
binding energy per particle=16 MeV, saturation density $\rho_0=.16 fm^{-3}$,
compressibility $K$=215 MeV and $m^*/m$=.67 at the fermi energy;
$u(\rho,p)$ gives the correct general behaviour of the real part
of the optical potential as a function of incident energy.  A
comparison of $u(\rho,p)$ with that derived from UV14+UVII potential
in cold nuclear matter can be found in \cite{Gale1}.  The specific
functional form of the momentum dependent part arises from the Fock
term of an Yukawa potential.   Mean fields given by eqs. (1)
and (2) have been widely 
tested for flow data \cite{Zhang} and give very good agreement.

To do a finite temperature calculation the following steps have to be
executed.  We need to find the occupation probability
\begin{eqnarray}
n[\epsilon(p)]=\frac{1}{e^{\beta[\epsilon(p)-\mu]}+1}
\end{eqnarray}
for a given temperature $1/\beta$ and density $\rho$.  If $\epsilon(p)$
were known {\it a priori}, this would merely entail finding the chemical
potential from
\begin{eqnarray}
\rho=\frac{16\pi }{h^3}\int_0^{\infty} p^2n[\epsilon(p)]dp
\end{eqnarray}
But the expression for $\epsilon(p)$ is
\begin{eqnarray}
\epsilon(p)=\frac{p^2}{2m}+A[\frac{\rho}{\rho_0}]+B[\frac{\rho}{\rho_0}]^
{\sigma}+R(\rho,p)
\end{eqnarray}
where at finite temperature
\begin{eqnarray}
R(\rho,p)=2\frac{C}{\rho_0}\frac{4}{h^3}\int d^3p'n[\epsilon(p')]\times
\frac{1}{1+[\frac{\vec p-\vec p'}{\Lambda}]^2}
\end{eqnarray}
Thus knowing $R(\rho,p)$ requires knowing $n[\epsilon(p')]$ 
already for all values of
$p'$.  This self-consistency condition can be fulfilled by an iterative
procedure (details can be found in \cite{Gale1}).

To calculate pressure, we use the thermodynamic identity $pV=-E+TS+\mu N$
which then gives
\begin{eqnarray}
p=a+b+c 
\end{eqnarray}
Here
\begin{eqnarray}
a=-v-\frac{16\pi}{h^3}\int_0^{\infty}dp\frac{p^4}{2m}n[\epsilon(p)]
\end{eqnarray}
where $v$ is given by eq. (1) and the second term is the contribution from
the kinetic energy.
\begin{eqnarray}
b=-T\frac{16\pi}{h^3}\int_0^{\infty}p^2[n\ln n+(1-n)\ln (1-n)]dp
\end{eqnarray}
The term $c$ is $\mu\rho$.  The free energy per particle is $-(a+b)/\rho$.
The test $p=0$ when the minimum of free energy is reached provides
a sensible test of numerical accuracy.

Fig.1 shows a plot of free energy per particle against temperature.  
There is no minimum in free energy at 12 MeV although at lower temperatures
shown in the figure, minima are seen.  The figure therefore suggests that
as the system approaches the temperature of 12 MeV, it will break up.
In Fig.2 we show the caloric curve.  The temperature flattens against
$e^*$, momentarily reaches zero slope (more about the slope in a later
section) and we reach the end of the model.  At higher temperatures
there is no minimum in free energy and so to continue a new principle has
to be postulated.  If now we use a freeze-out volume then we have reached
the more widely used approach for the caloric curve.

We have compared the caloric 
curve with the one computed with the low enrgy expansion of the
Fermi-gas model: $e^*=aT^2$.  As in \cite{Sobotka1}, with momentum
dependence, the caloric curve is initially above the Fermi-gas model
curve and meets the Fermi-gas curve at a later point.  The details
of the caloric curve depends upon $m^*/m$.  Normally one writes
$m^*/m=m_km_{\omega}$.  Sobotka et al. adopt a phenomenolgical expression
for $m_k$ and $m_{\omega}$ \cite{Mahaux}.  In the momentum dependent 
calculation done here $m_k$ is in but $m_{\omega}$ is not.  Further,
Sobotka et al. use a finite nucleus of mass 197 but our calculation
is for an infinite nucleus.  In spite of these differences there is
remarkable similarity between the calculated caloric curves.

In a previous report \cite{Dasgupta3} the finite temperature mean
field theory caloric curve of Fig.2 appears
but it does not continue after it
intersects the Fermi-gas model curve.  The temperature
interval in that calculation was not small enough and thus missed
the flattening of the caloric curve.  Sobotka \cite{Sobotka2} has pointed
out that the inclusion of $m_{\omega}$ further lifts $T$ at low $e^*$
and flattens it at high $e^*$.

\section{Isotherms, Maxwell construction and zeroes of pressure}
We now restrict ourselves to just density dependent
but momentum independent mean field.  There are two reasons for choosing
this: (a) calculations are much simpler and (b) by comparing with the 
caloric curve obtained above we will gain an understanding of how 
momentum dependence modifies the caloric curve.
This means the $C$ term in eqs. (1) and (2)
is put to zero and $A,B$ and $\sigma$ are readjusted to give desired values
of binding energy per nucleon, equilibrium density and compressibility.
The constants now are $A$=-356.8 MeV, $B$=303.9 MeV and $\sigma=$7/6.

Our objective is to obtain the caloric curve but we first look at
the well-known $p-\rho$ diagrams for constant temperature (isothermals)
and associated liquid-gas co-existence curves.
We discuss this by referring to Fig.3
where we have drawn isothermals corresponding to temperatures 6, 8,
10, 12, 14 and 15.64 MeV.  The extrema of free energy appear where
these isothermals intersect the $p$=0 line.  All isothermals will
reach $p$=0 at the uninteresting $\rho$=0 limit but the lower temperature 
isothermals intersect the $p$=0 line also at two other values of $\rho$.
Of these, the higher value of $\rho$ correspond to the minimum
of free energy
(this is the one that has concerned us above) and the lower value
of $\rho$ corresponds to a maximum of free energy and is of no
consequence to us in this section but will become very relevant
in the next section when we consider constant energy (rather than
constant temperature) solutions.  We have also drawn the co-existence line
following the procedure outlined in \cite{Reif}.  The high density
side of the co-existence line is of interest here.  The co-existence
line encloses the $p$=0 line (if not, the Maxwell construction
would lead to contradiction).  However, at low temperatures the value
of the density where the co-existence line intersects an isothermal
is very close to the value of the density where the isothermal intersects
the $p$=0 line.  For example, the value of the density where the 
co-existence line intersects the $T$=6 MeV line is 0.148 fm$^{-3}$.
The value of the density where this isothermal cuts the $p$=0 line
is very close but marginally less.  At $T$=12 MeV the density at
which the isothermal intersects the co-existence line is 0.111 fm$^{-3}$.
The $p$=0 line intersects this isothermal at a density which is only
0.011 fm$^{-3}$ less.  This means that with the usual interpretation
when the system has reached $p$=0 it is already in a mixed phase
but only a tiny fraction is in the gas phase.

The caloric curve in finite temperature simplified mean field model
(neglect of momentum dependence) is shown in Fig.4.  This curve is
below the curve obtained with simple Fermi-gas approximation.
Comparison with Fig.2 shows the effect of momentum dependence. 
Here also the system expands with temperature.  There is no minimum
in free energy beyond temperature $\approx$ 12.2 MeV.  Thus the 
caloric curve disappears afterwards unless a different prescription is
given for choosing an appropriate density.  Before the caloric curve 
disappears the slope of $T$ will reach zero which can be seen from
the arguments given in the next section.
\section{Constant energy mean field solutions: appearence of negative
specific heat}
We will now consider constant energy rather than constant temperature
solutions.  We use the density dependent but momentum independent
interaction of section IV.
We pick a $e^*$=excitation energy per particle.  At a
given density we guess a temperature and calculate what $e^*$ it generates
with equilibrium Fermi occupation factors.
The temperature is then varied till the correct $e^*$ is obtained.
Next we pick another density and again vary the temperature to obtain
the prescribed $e^*$.  Now a different temperature will be obtained.
The reader will notice that we are neglecting
the fluctuation of $E^*$ obtained
from the grand canonical calculation.  Since this is
an infinite matter calculation, this is justified.  At each density
we also calculate entropy per particle $s$ and the equilibrium
density is chosen from the condition 
$\frac{\partial s}{\partial\rho}|_{e^*}=0$ (and the second derivative is
positive). 
At this density the temperature can be can be defined by 
$\frac{\partial e^*}{\partial s}|_{\rho}=T$. (This definition gives
no difference from the temperature used in the
grand canonical calculation).
The pressure is defined by $p=\rho^2\frac{\partial e^*}{\partial\rho}|_{s}$.
Thus the caloric curve can be traced out and the pressure checked.
At the density where the entropy maximises, the pressure goes to zero.
For calculation of entropy one can follow the method described in 
\cite{Bohr}.  We define
$S=\ln \Omega(E^*)$ where $s=S/N$ and $e^*=E^*/N$ where $N$ is the number
of particles (taken to be arbitrarily large).  Because of Pauli principle,
$\Omega(E^*)$ is very difficult to calculate directly.  However it
can be obtained from Laplace inverse of the grand partition function
(which respects the Pauli principle)
in the saddle-point approximation.  This leads to
\begin{eqnarray}
S=(E^*-\mu N)/T+\ln Z_{gr}(\mu,T)
\end{eqnarray}
Here $T$ is the temperature and $\mu$ the chemical potential
which give the prescribed $E^*$ and $\rho$ when
evaluated in the grand canonical ensemble.  
In eq. (10), we have neglected a pre-factor
as we only need $s=S/N$ and in the large $N$ limit the pre-factor gives
zero contribution.

We mention here that instead of eq.(10) one can also use the more
well-known formula for entropy:
\begin{eqnarray}
S=-\sum_i[n_i\ln n_i+(1-n_i)\ln (1-n_i)]
\end{eqnarray}
Both give the same result.  Similarly instead of the microcanonical
expression for pressure one can use the the grand canonical
expression for pressure (i.e., eq.7 with $C$ of eq.1 set to 0).
 
The $p-\rho$ curves for constant $e^*$ are shown in Fig.5.  Compared
to the isothermal $p-\rho$ curves, there are two huge differences.
One is that for $e^*$ below 16 MeV (which is the binding energy
per particle at saturation density 
in mean field theory) the constant energy $p-\rho$ curves terminate
in the low density side.  This is when the stretching energy
per particle equals $e^*$.  The other difference is that below
$e^*$=16 MeV, these curves intersect the $p$=0 line at only
one point.  The isothermal $p-\rho$ diagrams (Fig.3) intersect the $p$=0
line at two points (we are ignoring the uninteresting case of 
$\rho$=0).  All the zeroes of $p$ in Fig.3 must also appear in Fig.5.
How do they map? The zeroes of pressure in Fig.5 are all maxima in 
$s$ if one moves along the constant energy contour but not all of them
correspond to minima of $f$ if one moves along the constant temperature
contour.

For our specific case, the zeroes of pressure at $e^*$=6, 8, 10
and 12 MeV (Fig. 5) correspond to minima of free energy $f$ (Fig.3).
The zeroes of pressure for $e^*$=14 and 15 MeV (Fig.5) correspond to maxima
of $f$.  Thus all the minima in $f$ of Fig.3 are bunched into
zeroes of $p$ below $e^*\approx$=13.6 MeV and all the maxima of $f$ in Fig.3
are bunched into zeroes of $p$ with $e^*\approx$13.6 to 16 MeV.  It then
follows that in the caloric curve $T$ increases with $e^*$ and must
flatten out (this is the end of the curve in the constant temperature
model).  In the constant energy model, it will continue beyond finally
terminating at $e^*$=16 MeV.  But when the caloric curve extends beyond
the finite temperature model, the slope of $T$ against $e^*$ must be negative.
This is a physically unreasonable result since in 
the grand canonical calculation we are
dealing with an arbitrarily large system.  This reflects the inadequacy
of the mean field model.  Indeed in terms of Fig.3 we are in the
negative compressibility zone.

We note in passing that negative specific heat also appears in the figures
of \cite{Sobotka1}.  This is due to inadequacy of the model employed
\cite{Sobotka2}.

\section{Discussion}
The variational principle model for the nuclear caloric curve is also
capable of producing a maximum in the specific heat.  This maximum is
gently approached as opposed to that in the thermodynamic model where 
it is a very sharp
peak \cite{Dasgupta1,Dasgupta2} in the nuclear matter limit.
The model likely has validity for
proton induced reactions as opposed to central collisions of heavy ions.
To be of greater use one needs to extend the model to higher energy.
It is not clear how this is to be done.  We also notice that the 
interesting part of the caloric curve is at a rather high value
of temperature (and excitation energy) compared to what we encounter
in experiments.  This may be a shortcoming of the mean field model or
the infinite matter limit or both.

\section{Acknowledgement}
I thank Lee Sobotka for many communications.  This work is supported
in part by the Natural Sciences and Engineering Research Council of Canada
and in part by the Quebec Department of Education.

\begin{figure}
\epsfxsize=5.5in
\epsfysize=7.0in
\centerline{\epsffile{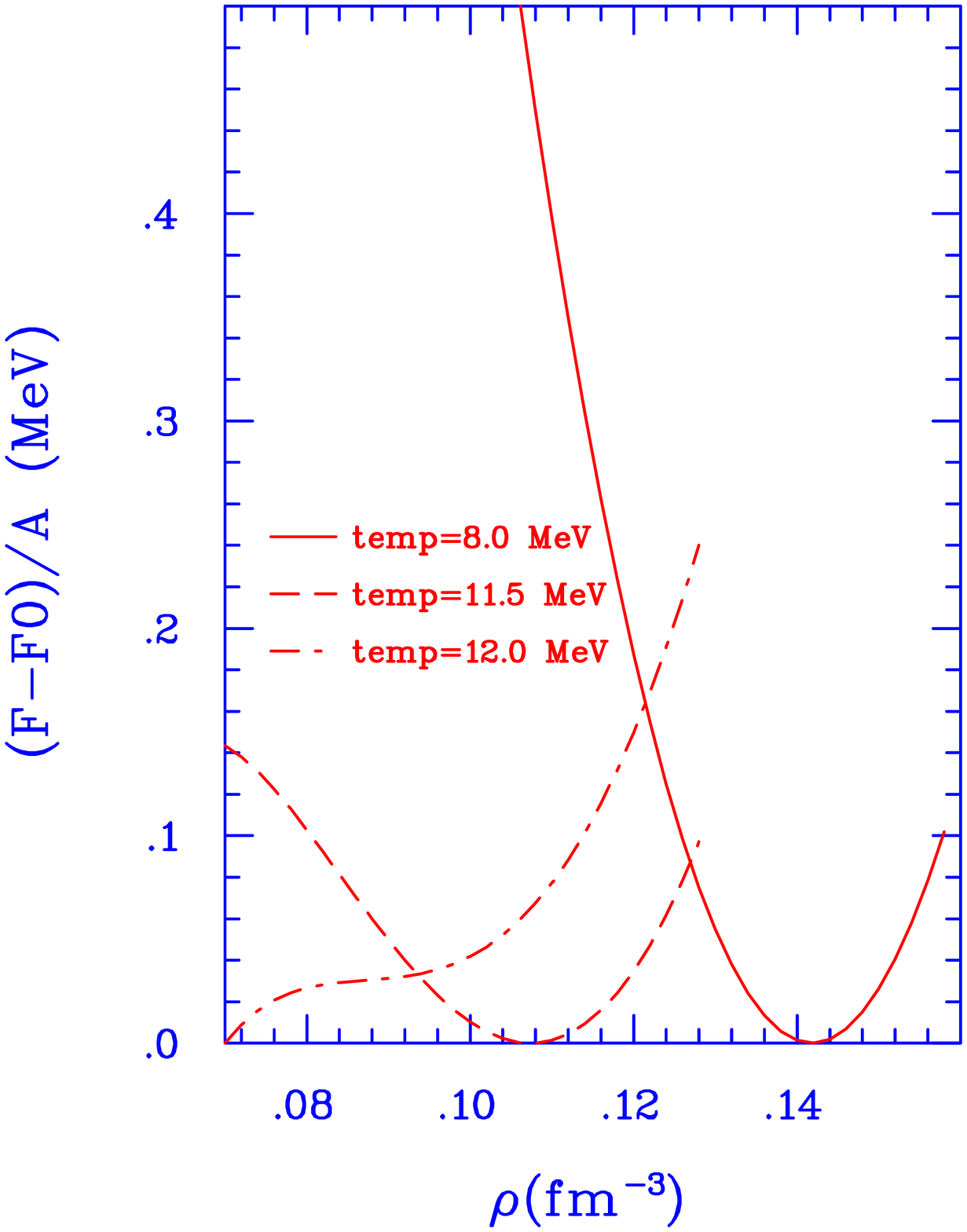}}
\vskip 0.8 true cm
\caption{The free energy per particle as function of density at different
temperatures.  This is plotted as a difference from the minimum value of
F in the frame. At temperature 12 MeV there is no minimum and the system
will roll down to lower density.  For this temperature  already at 
lower values of $\rho$ in the figure, the 
derivative $\frac{\partial p}{\partial\rho}$ is negative, i.e.,
the system has entered a region of mechanical instability.} 
\end{figure}

\begin{figure}
\epsfxsize=4.5in
\epsfysize=6.0in
\centerline{\epsffile{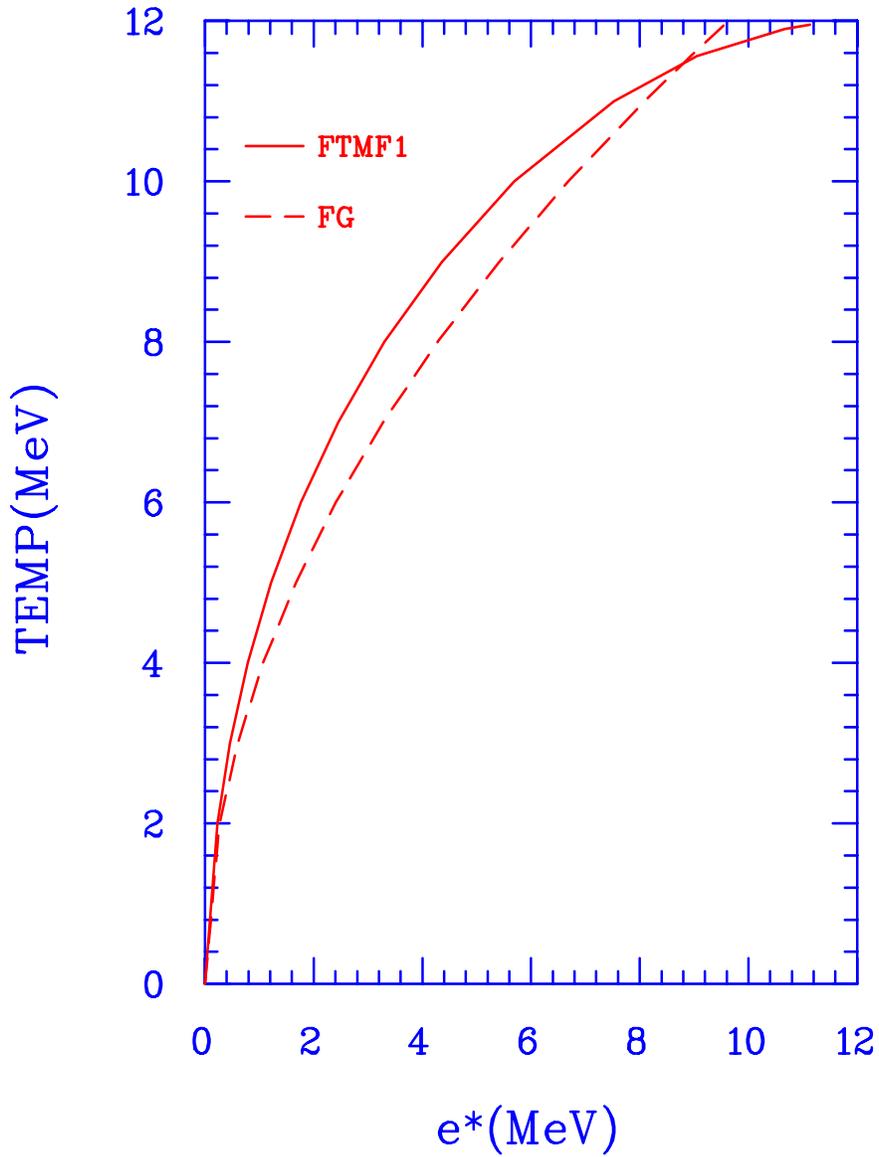}}
\vskip 0.8 true cm
\caption{Temperature against e* in finite temperature
mean field model with momentum
dependence (FTMF1) and in Fermi gas model (FG).  As in ref. 6, the curve with
momentum dependence lies above the Fermi gas curve till they meet
(around temperature 11.5 MeV).  Beyond the FTMF1 curve flattens.
Just below temperature 12 MeV the curve stops.  The nucleus will break
up into many pieces at higher temperature.}
\end{figure}

\begin{figure}
\epsfxsize=4.5in
\epsfysize=6.0in
\centerline{\epsffile{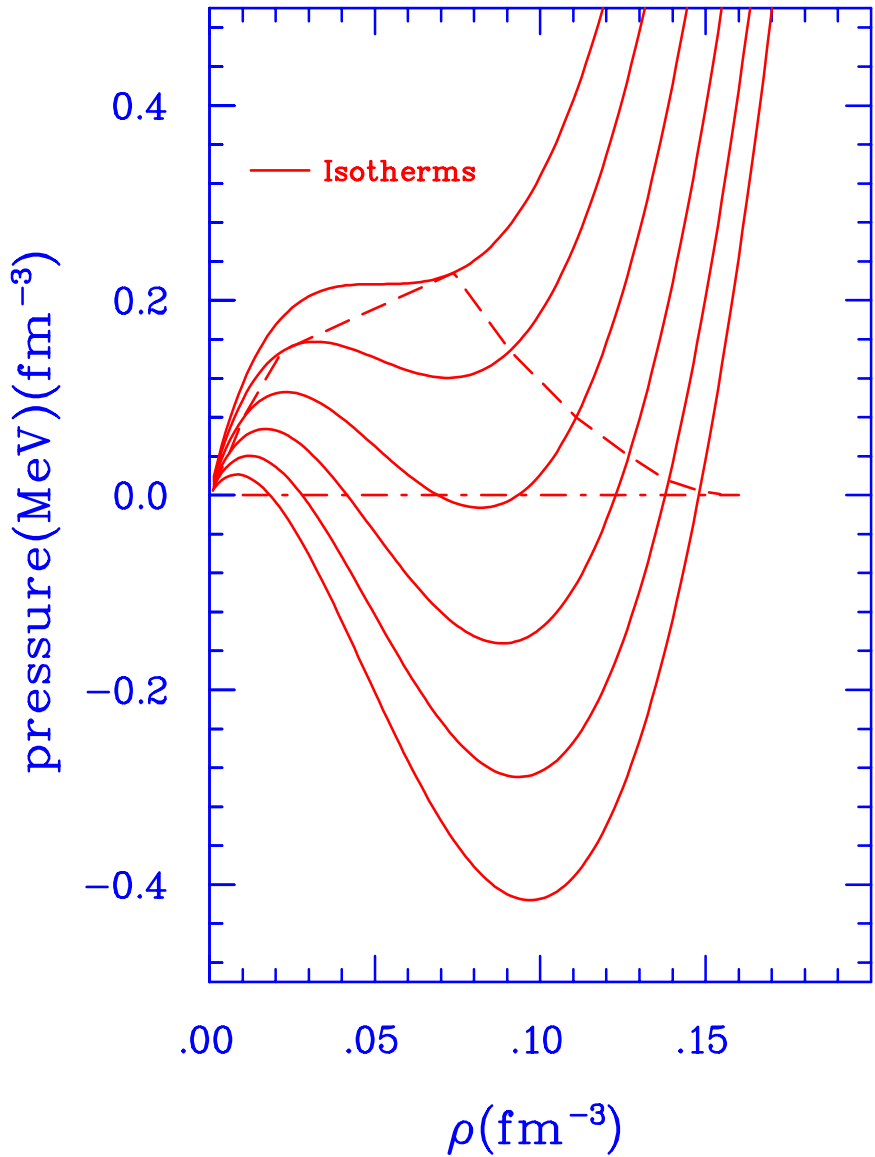}}
\vskip 0.8 true cm
\caption{Isotherms at temperatures 6, 8, 10, 12, 14 and 15.64 MeV in
mean field model with density dependence but no momentum dependence.
The dashed line is the co-existence line.  Also shown is the $p$=0
line.  The intersections of this line with the isotherms give the
the minima (in the high density side) and maxima (in the low density side)
of free energy.}
\end{figure}

\begin{figure}
\epsfxsize=4.5in
\epsfysize=6.0in
\centerline{\epsffile{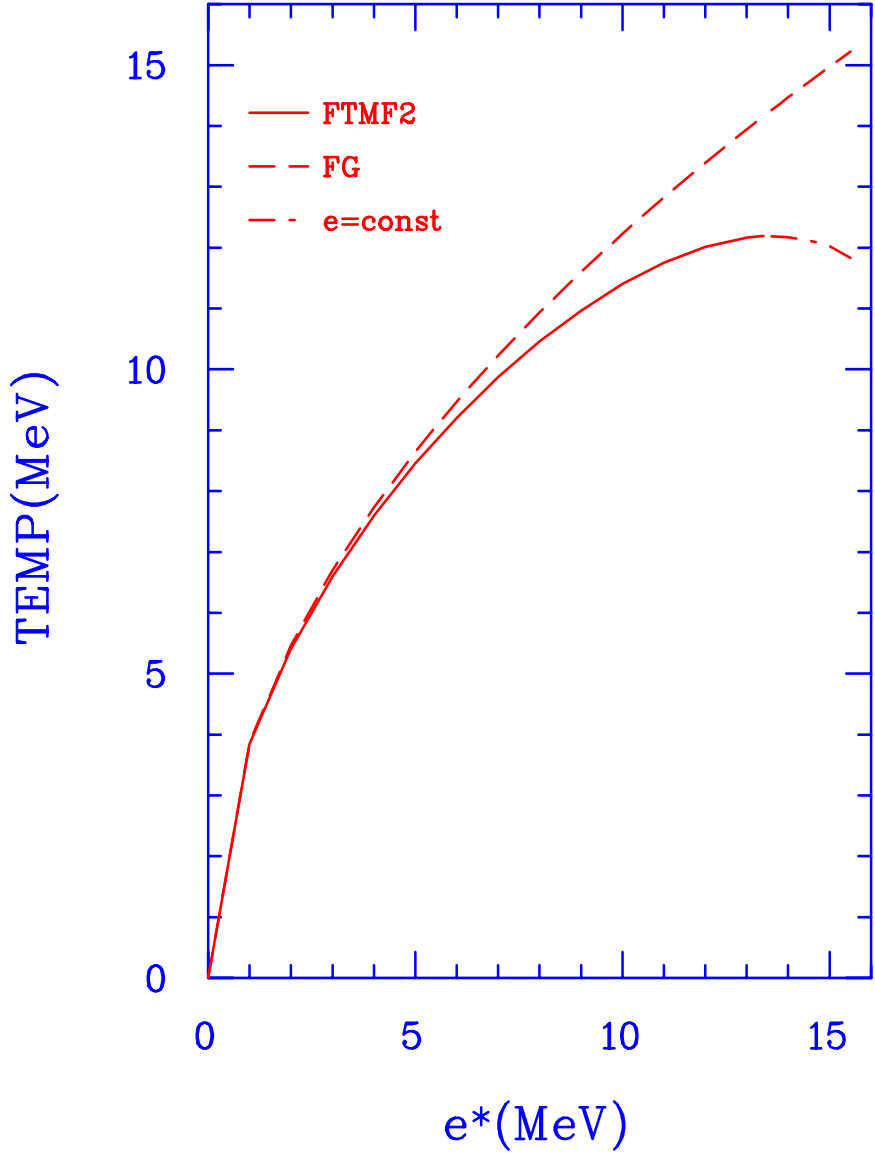}}
\vskip 0.8 true cm
\caption{The caloric curve according to finite temperature mean field
theory (solid curve), according to finite excitation mean field theory
and $e^*=aT^2$ Fermi-gas approximation (dashed curve).  The finite temperature
and finite excitation answers are the same upto $e^*\approx$13.6 MeV.
The finite temperature model stops here with zero slope for $T$.  The
finite excitation model continues beyond but with an unphysical negative
slope for $T$.  This part belongs to the negative compressibility zone
of Fig.3.}
\end{figure}

\begin{figure}
\epsfxsize=4.5in
\epsfysize=6.0in
\centerline{\epsffile{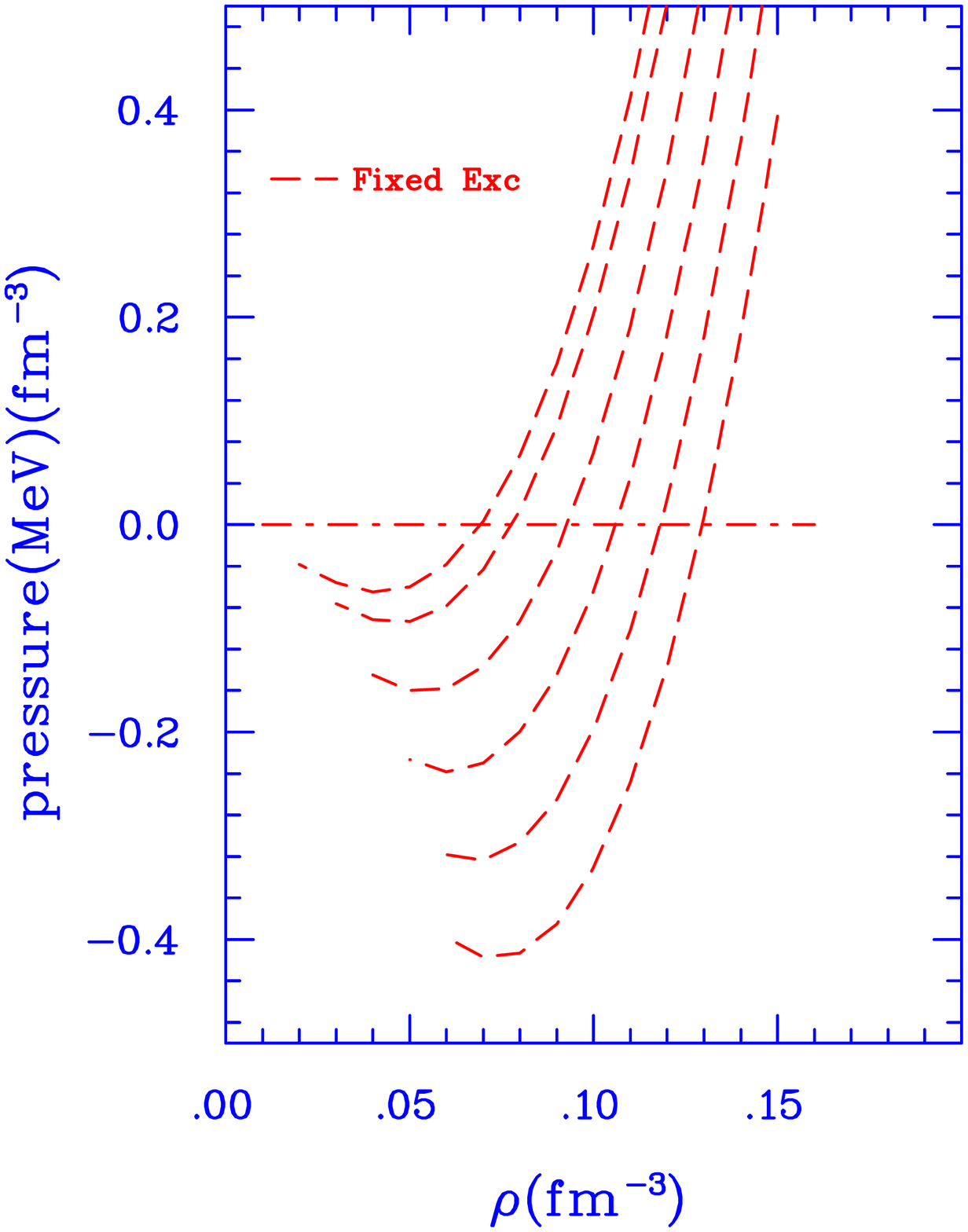}}
\vskip 0.8 true cm
\caption{Pressure against $\rho$ curves for fixed excitation energies
which in this case are, starting from the lowest, for $e^*$=
6, 8, 10, 12, 14 and 15 MeV.  The curves terminate on the low density side
when the energy cost of 
expansion equals $e^*$.  The curves intersect the $p$=0 line at
only one point unlike the case for isotherms (Fig.3).}
\end{figure}


\begin{references}

\bibitem{Dasgupta1} S. Das Gupta, A. Z. Mekjian and M. B. Tsang, Advances
in Nuclear Physics, vol.26, 91 (2001)
 
\bibitem{Bondorf1} J. P. Bondorf, R. Donangelo, I. M. Mishustin, and
H. Schulz, Nucl. Phys. {\bf A444},321 (1985)

\bibitem{Pochodzalla} J. Pochodzalla et al., Phys. Rev. Lett. {\bf 75},
1040 (1995)

\bibitem{Dasgupta2} C. B. Das, S. Das Gupta, W. G. Lynch, A. Z. Mekjian,
and M. B. Tsang, Phys. Rep. {\bf 406}, 1, (2005)

\bibitem{Bertsch} G. F. Bertsch and S. Das Gupta, Phys. Rep. {\bf 160},
189 (1988)

\bibitem{Sobotka1} L. G. Sobotka, R. J. Charity, J. Toke and W. U.
Schroder, Phys. Rev. Lett. {\bf93 },132702 (2004) 

\bibitem{Toke} J. Toke, J. Lu and W. Udo Schroder, Phys. Rev. C{bf 67},
034609 (2003)

\bibitem{March} N. H. March, W. H. Young and S. Sampanther, {\it
The Many-Body problem in Quantum Mechanics},(Cambridge University
Press, Cambridge, 1967)p. 243

\bibitem{De} J. N. De, S. Das Gupta, S. Shlomo, and S. K. Samaddar,
Phys. Rev C{\bf 55},R1641 (1997)

\bibitem{Welke} G. M. Welke, M. Prakash, T.T. S. Kuo, S. Das Gupta,
and C. Gale, Phys. Rev. C{\bf 38}, 2101 (1988)

\bibitem{Gale1} C. Gale, G. M. Welke, M. Prakash, S. J. Lee and 
S. Das Gupta, Phys. Rev. C{\bf 41}, 1545 (1990)

\bibitem{Zhang} J. Zhang, S. Das Gupta and C. Gale, Phys. Rev. C{\bf 50},
1617 (1994)

\bibitem{Dasgupta3} S. Das Gupta, nucl-th/0412057 v1,15 Dec.,(2004)

\bibitem{Sobotka2} L. Sobotka, private communication.
\bibitem{Mahaux} C. Mahaux, P. F. Bortignon, R. A. Broglia and C. H. Dasso,
Phys. Rep. {\bf 120}, 1 (1985)

\bibitem{Reif} F. Reif, {\it Fundamentals of statistical and thermal 
physics}(McGraw Hill, New York, 1965)Chapter 8.

\bibitem{Bohr} A. Bohr and B. R. Mottelson, {\it Nuclear Structure, Vol.1}
(W. A. Benjamin Inc, New York, 1969)Appendix 2B

\end{references}
\end{document}